# An inverse and analytic lens design method

Yang Lu,[1] Vasudevan Lakshminarayanan,[1, 2]

[1] School of Optometry and Vision Science, University of Waterloo, 200 University Avenue West, Waterloo, ON, N2V 2V8
[2] Departments of Physics and Electrical Engineering and Computer Science, University of Wterloo,

Traditional lens design is a numerical and forward process based on ray tracing and aberration theory. This method has limitations because the initial configuration of the lens has to be specified and the aberrations of the lenses have to be considered. This letter is an initial attempt to investigate an analytic and inverse lens design method, called Lagrange, to overcome these barriers. Lagrange method tries to build differential equations in term of the system parameters and the system output and input (object and image). The generalized Snell's law in three-dimensional space and the normal of a surface in fundamental differential geometry are applied. Based on the Lagrange method, equations for a single surface system are derived, which can perfectly image a point object.
.

## 1. Introduction

Traditional lens design is based on ray tracing and aberration theory, which consider the lens system as being rotationally symmetric around an optical axis.[1]. Automatic lens design method, i.e. optimization, is widely used by optical designers[2]. Optimization is a forward and two-dimensional design process, in which we derive refracted rays from the given incident ray and the nature of the surface. The performance of the lens is confined by the initial shape parameters of the system, such as refractive index, radius and thickness. In order to overcome this barrier, an inverse and analytic design method is introduced. We have named this as Lagrange. Lagrange can find the solution of an unknown surface by the given incident ray and refracted ray. This is a three-dimensional lens design concept, which combines the normal of a surface in differential geometry and Snell's law in vector form. The rest of the letter is organized as follows. First, spherical coordinates where the lens system is constructed are introduced. Second, we introduce Snell's law in vector form in three-dimensional space. Finally, we derive the partial differential equations for a single surface system in spherical coordinates.

## 2. Mathematical Preliminaries

Spherical coordinates are used to simplify the equation derivations. The transformation between Cartesian coordinates, $(x,y,z)$, and spherical coordinates, $(r,\theta,\varphi)$, is: (see figure 1), [3]:

$$x = r\,sin\theta\,cos\varphi$$
$$y = r\,sin\theta\,sin\varphi \quad (1)$$
$$z = r\,cos\theta$$

We define a position vector as:

$$\vec{r} = x\,i + y\,j + z\,k \quad (2)$$

where $i, j$ and $k$ are Cartesian coordinate unit vectors (basis) corresponding to $x$, $y$ and $z$ axis respectively. Based on this transformation, three unit vectors (basis) in spherical coordinates are simply:

$$\vec{e}_r = \frac{\partial\vec{r}/\partial r}{|\partial\vec{r}/\partial r|} = sin\theta\,cos\varphi\,i + sin\theta\,sin\varphi\,j + cos\theta\,k$$

$$\vec{e}_\theta = \frac{\partial\vec{r}/\partial\theta}{|\partial\vec{r}/\partial\theta|} = cos\theta\,cos\varphi\,i + cos\theta\,sin\varphi\,j - sin\theta\,k \quad (3)$$

$$\vec{e}_\varphi = \frac{\partial\vec{r}/\partial\varphi}{|\partial\vec{r}/\partial\varphi|} = -sin\varphi\,i + cos\varphi\,j$$

where $\theta$ is the angle down from the north pole, and $\varphi$ is the angle around the equator, as shown in figure 1. These transformation equations are the basis of the derivations introduced in the following sections.

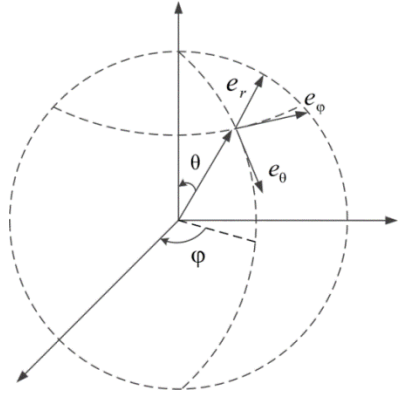

Figure 1 Spherical Coordinates

Figure 2 illustrates the definition of the normal of a surface. In spherical coordinates, an arbitrary surface can be expressed as:

$$\vec{f} = f(\theta,\varphi)\,\vec{e}_r \quad (4)$$

where $\vec{f}$ is the position vector for the arbitrary point on the surface. $f(\theta,\varphi)$ is the function of $\theta$ and $\varphi$, which represents the magnitude of $\vec{f}$. Notice that $\vec{e}_r$ is also the function of $\theta$ and $\varphi$ in terms of the equation 3. We define the normal of the surface as the composite function of the surface function, $f$, shown as:

$$\vec{N} = \vec{g}(f) \quad (5)$$

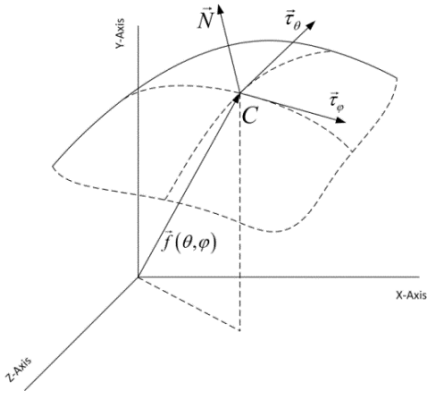

Figure 2 The normal of a surface

For any arbitrary point on the surface, the normal of the surface is expressed as: [4]

$$\vec{N} = \vec{g}(f) = \frac{\partial \vec{f}}{\partial \theta} \times \frac{\partial \vec{f}}{\partial \varphi} = f^2 \sin\theta \, \vec{e}_r - f \sin\theta \, p \, \vec{e}_\theta - f \, q \, \vec{e}_\varphi$$

$$p = \frac{\partial f(\theta,\varphi)}{\partial \theta}; q = \frac{\partial f(\theta,\varphi)}{\partial \varphi} \qquad (6)$$

The function of the normal of the surface is an important concept in the Lagrange design method.

### 3. Snell's law in three-dimensional space

Here, Snell's law is generalized to vector form so that the lens can de designed in three-dimensional space.

In two-dimensional space, Snell's law is expressed as:

$$n_1 \sin\alpha_1 = n_2 \sin\alpha_2 \qquad (7)$$

where $\alpha_1$ is the incident angle. $\alpha_2$ is the refracted angle. $n_i$ ($i = 1,2$) is the index. However, in three-dimensional space, Snell's law is presented as a vector form which is more general than that given by equation 7. Assume that $\vec{p}_1$ is the unit direction vector of the incident ray. $\vec{p}_2$ is the unit direction vector of the refracted ray. $\vec{u}$ is the unit direction vector of the normal, as shown in Figure 3. Because $\vec{p}_1$, $\vec{p}_2$, and $\vec{u}$ are in the same plane and satisfy Snell's law, we can have $\vec{u}$ as a linear combination of $\vec{p}_1$ and $\vec{p}_2$, shown as:

$$U \vec{u} = n_2 \vec{p}_2 - n_1 \vec{p}_1 \qquad (8)$$

where $|\vec{u}| = |\vec{p}_2| = |\vec{p}_1| = 1$.

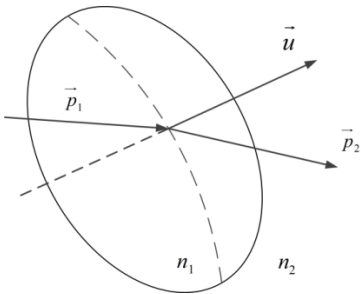

Figure 3 Snell's law in vector form

The surface normal function in spherical coordinates and the Snell's law in vector form are two important concepts for the Lagrange method. By combining these two concepts, we can derive the differential equations in the lens system.

### 4. Fundamental Partial Differential Equation for Lens system

For simplicity, we start from a single surface imaging system. This system is an interface between two different media, which can perfectly image a point object. Figure 4 illustrates its layout. Point $s$ is an object on Z axis. $s'$ is the image. $\vec{s}$ represents the position vector of the point $s$. $\vec{R}_1$ represents an incident ray vector, and $\vec{R}_2$ represents the refracted ray vector. $\vec{u}$ represents the normal direction, which is a unit vector. $\vec{f}$ represents the surface function.

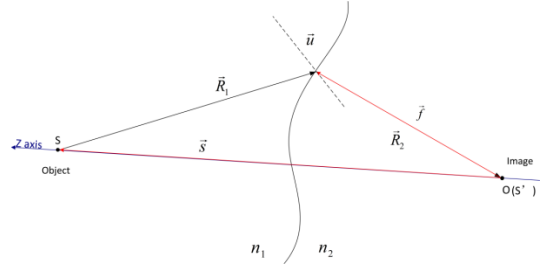

Figure 4 The single surface system

The core idea of the Lagrange method is that the normal of the surface and the normal in Snell's law have the same direction. From equation 6, we have the surface normal function, $\vec{N}$. If we define the normal in Snell's law as $\vec{U}$. According to equation 8, $\vec{U} = U \vec{u}$, and $U = |\vec{U}|$. Therefore, we have:

$$\frac{\vec{N}}{|\vec{N}|} = \pm \frac{\vec{U}}{|\vec{U}|} \qquad (9)$$

If we substitute the equation 6 into 9, we finally have:

$$(f \sin\theta \, p \, U_1 + f^2 \sin\theta \, U_2)^2 + (U_1^2 + U_2^2) f^2 q^2 = 0 \qquad (10)$$

where $U_1$, $U_2$ and $U_3$ are the three components along the $\vec{e}_r$, $\vec{e}_\theta$ and $\vec{e}_\varphi$ directions respectively, and $U_3$ is zero because we define that both object and image are on z axis. This equation is correct if and only if $f \sin\theta \, p \, U_1 + f^2 \sin\theta \, U_2 = 0$ and $q^2 = 0$. Therefore, we have the differential equations of the single surface system:

$$\begin{cases} p \, U_1 + U_2 \, f = 0 \\ q = 0 \end{cases}$$

$$p = \frac{\partial f}{\partial \theta}; q = \frac{\partial f}{\partial \varphi} \qquad (11)$$

### 5. Conclusion

We have defined an inverse and analytic lens design method, Lagrange, which is different from traditional forward design method. Lagrange method is based on the idea that the normal of the surface has the same direction as the normal in Snell's law. In terms of these two concepts, differential equations of a single surface system which perfectly image a point are derived. The surface solution can be obtained by solving the equations. We have used this method to derive lens systems which have zero aberrations and these results are presented elsewhere. [5]